\documentstyle[preprint,aps]{revtex}
\input epsf.tex
\def\DESepsf(#1 width #2){\epsfxsize=#2 \epsfbox{#1}}
\begin{document}
\preprint{\vbox{\hbox{OITS-585}}}
\draft
\title{CP Asymmetry In Neutral $B$ System At Symmetric Colliders\footnote{Work
supported in part by the Department of Energy Grant No.
DE-FG06-85ER40224.}}
\author{N.G. Deshpande, and Xiao-Gang He}
\address{Institute of Theoretical Science\\
University of Oregon\\
Eugene, OR 97403-5203, USA}

\date{September 1995}
\maketitle
\begin{abstract}
Contrary to the conventional belief, time integrated asymmetry are
measurable in selected final states in the neutral $B$ system at
symmetric $e^+e^-$ colliders. They occur due to the interplay of weak and
strong phases of two different amplitudes in addition to
the $B^0 - \bar B^0$ mixing. Observation of these asymmetries
would be evidence for direct CP violation in the decay amplitudes.
\end{abstract}
\pacs{}
CP violation is one of the few remaining unresolved myseries in
particle physics. The explanation in the Standard Model based on
Cabibbo-Kobayashi-Maskawa (CKM) matrix is still not established.
Although there is no conflict between the observation of CP violation
in the K-system and theory\cite{1}, intriguing hints of other plausible
explanations emerge from astrophysical considerartion of baryon to
photon ratio in the universe\cite{2}.
Models based on additional Higgs bosons or gauge bosons can equally well
explain the existing data\cite{3}. It is for this reason that exploration of CP
vioaltion in the $B$ system is so crucial. The $B$ system offers several
final states that provide a rich source for the study of this
phenomena\cite{stone}.
The principle techniques at electron-positron colliders that will be used
involves: (a) measurement of particle-antiparticle partial rate asymmetries,
and (b) rate aysmmetries in the neutral $B$ system with a lepton tag for one
of the $B$ mesons and the decay of the other $B$ into a CP eigenstate
f (e.g. f - $\psi K_S$)\cite{4},

\begin{eqnarray}
Asy = {(l^+ f) - (l^- f)\over (l^+f) + (l^-f)}\;.
\end{eqnarray}

The rationale for building an asymmetric electron-position collider
arises from the well known observation that time integrated asymmetries
arising from $B^0-\bar B^0$ production in $C = -1$ state are washed out
due to quantum coherence of the initial state. On closer examination this
is exactly correct only when the amplitude for the process $B^0 \rightarrow f$
has a single weak phase. As we show below, when the amplitude is a mixture of
two terms with different weak and strong phases, the asymmetry though diluted,
still remains.

Consider the deay of the coherent $B^0\bar B^0$ state produced in $C = -1$
state
as in the decay of $\Upsilon(4s)$. If $t_1$ and $t_2$ denote decay times of the
states that were pure $B^0$ and $\bar B^0$ at time zero, and $f_a$ is a flavor
specific decay like $l^-\bar \nu D^*$ that tags pure $\bar
B^0$ state, while $f_b$ is a CP eigenstate (e.g. $\psi K_S$, $\pi^+\pi^-$,
$\pi^0\pi^0$ ...), then the rate is given by\cite{4}
\begin{eqnarray}
Rate (B^0(t_1)\bar B^0(t_2)\rightarrow f_a f_b)
&\sim& e^{-\Gamma(t_1+t_2)} \{ [1+cos(\Delta m(t_1-t_2)]|A(f_b)
\bar A(f_a)|^2\nonumber\\
&+& [1 - cos(\Delta m(t_1-t_2))]|{q\over p}|^2 |\bar A(f_b)\bar A(f_a)|^2
\nonumber\\
&-& 2 sin\Delta m (t_1-t_2)|A(f_b)\bar A(f_a)|^2 Im\left (
{q\over p}{\bar A(f_b) \over A(f_b)}\right )\}
\end{eqnarray}
where $\Delta m$ is the mass difference between the heavier and ligher neutral
$B$ mesons, $A$ and $\bar A$ are amplitudes for $B^0\rightarrow f$ and
$\bar B^0\rightarrow f$ respectively,
and $q$ and $p$ are complex parameters defining $B^0$ and $\bar B^0$ mixing.
The rate for $B^0(t_1)\bar B^0(t_2)\rightarrow \bar f_a f_b$ is given by
similar
expression with replacement
\begin{eqnarray}
A(f_b) \rightarrow \bar A(f_b)\;,\nonumber\\
\bar A(f_a)\rightarrow A(\bar f_a)\;.
\end{eqnarray}
For the leptonic mode which involves a single weak phase, we have
\begin{eqnarray}
|\bar A(f_a)| = |A(\bar f_a)|\;.
\end{eqnarray}
In the $B$ system $|q/p| = 1$ to a very good approximation. If there is a
single
amplitude contributing to the decay $B^0\rightarrow f_b$, we have $|A(f_b)| =
|\bar A(f_b)|$ and the cosine term drops off. The asymmetry then is
proportional
to $\mbox{sin}(\Delta m(t_1-t_2))$. This term vanishes when integrated over
$t_1$, $t_2$
from zero to infinity as would be the case for a symmetric collider. Consider
now
the case where the contribution to the decay $B^0\rightarrow f_b$ contains two
contributions with different weak and strong phases. This situation arises when
a process has both tree and penguin contributions. We can write in generality
\begin{eqnarray}
A(B^0\rightarrow f_b) = Te^{i(\delta_w+\delta_s)} + P\;,
\end{eqnarray}
where $T$ and $P$ stand for the tree and penguin contributions, $\delta_w$
and $\delta_s$ are the weak and the strong relative phase between the tree and
penguin
amplitudes. We can now take T and P to be real. For the antiparticle amplitude
we have
\begin{eqnarray}
\bar A(\bar B^0\rightarrow f_b) = Te^{i(-\delta_w+\delta_s)} + P\;.
\end{eqnarray}

Time integrated rate for $B^0\bar B^0\rightarrow f_a f_b$ is now given by
\begin{eqnarray}
Rate \sim |\bar A(f_a)|^2\left[{|A(f_b)|^2 +|\bar A(f_b)|^2\over \Gamma^2}
+ {|A(f_b)|^2 - |\bar A(f_b)|^2\over \Gamma^2 +(\Delta m)^2}\right ]\;.
\end{eqnarray}

The asymmetry is given by
\begin{eqnarray}
Asy &=& {\Gamma^2\over \Gamma^2 + (\Delta m)^2}
{|A(f_b)|^2 - |\bar A(f_b)|^2\over |A(f_b)|^2 + |\bar A(f_b)|^2}\nonumber\\
&=& - X_d {2TP\mbox{sin}\delta_s \mbox{sin}\delta_w \over T^2+P^2 + 2 TP
\mbox{cos}\delta_s \mbox{cos}\delta_w}\;,
\end{eqnarray}
where $X_d = \Gamma^2/(\Gamma^2+(\Delta m)^2 ) \approx 0.5$\cite{5}, which is
the dilution factor.
It is important to note that the asymmetry arises from the weak phase in the
direct amplitude. Thus, in a superweak type model, this asymmetry would be
zero.
In the following  we consider three examples:
a) $f_b = \pi^0\pi^0$, b) $\bar B^0\rightarrow \pi^+\pi^-$, and
c) $f_b = \pi^0 K_S$.

In the SM the decay amplitudes for $\bar B^0\rightarrow \pi^0\pi^0$,
$\bar B^0\rightarrow \pi^+\pi^-$, and
$\bar B^0\rightarrow \pi^0 K_S$ are generated by the following effective
Hamiltonian:
\begin{eqnarray}
H_{eff}^q &=& {G_F\over \sqrt{2}}[V_{ub}V^*_{uq}(c_1O_1^q + c_2 O_2^q) -
\sum_{i=3}^{10}(V_{ub}V^*_{uq} c_i^u\nonumber\\
&+&V_{cb}V^*_{cq} c_i^c
+V_{tb}V^*_{tq} c_i^t)
O_i^q] +H.C.\;,
\end{eqnarray}
where the
superscript f in $c_i^f$
indicates the loop contribution from f quark, and $O_i^q$ are
defined as
\begin{eqnarray}
O_1^q &=& \bar q_\alpha \gamma_\mu Lu_\beta\bar
u_\beta\gamma^\mu Lb_\alpha\;,\;\;\;\;\;\;O_2^q =\bar q
\gamma_\mu L u\bar
u\gamma^\mu L b\;,\nonumber\\
O_{3,5}^q &=&\bar q \gamma_\mu L b
\bar q' \gamma_\mu L(R) q'\;,\;\;\;\;\;\;\;O_{4,6}^q = \bar q_\alpha
\gamma_\mu Lb_\beta
\bar q'_\beta \gamma_\mu L(R) q'_\alpha\;,\\
O_{7,9}^q &=& {3\over 2}\bar q \gamma_\mu L b  e_{q'}\bar q'
\gamma^\mu R(L)q'\;,\;O_{8,10}^q = {3\over 2}\bar q_\alpha
\gamma_\mu L b_\beta
e_{q'}\bar q'_\beta \gamma_\mu R(L) q'_\alpha\;,\nonumber
\end{eqnarray}
where $R(L) = 1 +(-)\gamma_5$,
and $q'$ is summed over u, d, and s. For $\Delta S = 0$ processes,
$q = d$, and for $\Delta S = 1$ processes, $q =s$. $O_{2}$, $O_1$ are the tree
level and QCD corrected operators. $O_{3-6}$ are the strong gluon induced
penguin operators, and operators $O_{7-10}$ are due to $\gamma$ and Z exchange,
and ``box'' diagrams at loop level. The Wilson coefficients $c_i^f$ are defined
at the scale of $\mu \approx
m_b$ which have been evaluated to the next-to-leading order in QCD\cite{6,7}.
We give these coefficients below for $m_t = 176$ GeV, $\alpha_s(m_Z) = 0.117$,
and $\mu = m_b = 5$ GeV\cite{7},
\begin{eqnarray}
c_1 &=& -0.307\;,\;\; c_2 = 1.147\;,\;\;
c^t_3 =0.017\;,\;\; c^t_4 =-0.037\;,\;\;
c^t_5 =0.010\;,
 c^t_6 =-0.045\;,\nonumber\\
c^t_7 &=&-1.24\times 10^{-5}\;,\;\; c_8^t = 3.77\times 10^{-4}\;,\;\;
c_9^t =-0.010\;,\;\; c_{10}^t =2.06\times 10^{-3}\;, \nonumber\\
c_{3,5}^{u,c} &=& -c_{4,6}^{u,c}/3 = P^c_s/3\;,\;\;
c_{7,9}^{u,c} = P^{u,c}_e\;,\;\; c_{8,10}^{u,c} = 0
\end{eqnarray}
where $c^t_i$ are the regularization scheme independent WC's obtained in Ref.
\cite{7}.
The leading contributions to $P^i_{s,e}$ are given by:
 $P^i_s = (\alpha_s/8\pi)\bar c_2 (10/9 +G(m_i,\mu,q^2))$ and
$P^i_e = (\alpha_{em}/9\pi)(3\bar c_1+\bar c_2) (10/9 + G(m_i,\mu,q^2))$.
The function
$G(m,\mu,q^2)$ is give by
\begin{eqnarray}
G(m,\mu,q^2) = 4\int^1_0 x(1-x) \mbox{d}x \mbox{ln}{m^2-x(1-x)q^2\over
\mu^2}\;.
\end{eqnarray}

Using the unitarity property of the CKM matrix, we can eliminate the term
proportional to $V_{cb}V^*_{cq}$ in the effective Hamiltonian. The $B$
decay amplitude due to the complex
effective Hamiltonian displayed above can be parametrized, without  loss of
generality, as
\begin{eqnarray}
<final\;state|H_{eff}^q|B> = V_{ub}V^*_{uq} T_q + V_{tb}V^*_{tq}P_q\;,
\end{eqnarray}
where $T_q$ contains the $tree\; contributions$ and $penguin\; contributions$
due to
u and c internal quarks, while $P_q$ only contains $penguin\; contributions$
from internal c and t quarks.

To obtain exclusive decay amplitudes, we need to calculate relevant hadronic
matrix elements.  Since no reliable calculational tool exists for two body
modes,
we shall use factorization
approximation to get an idea of the size of asymmetry Asy.
The numerical numbers obtained should be viewed as an order of magnitude
estimates.
The important message is that direct CP violations in decay amplitudes are
detectable.
Measurements of rate asymmetries at symmetric colliders will provide useful
information about CP violation. In the factorization approximation,
we have\cite{8}
\begin{eqnarray}
T_d(\pi^0\pi^0) &=& i{G_F\over \sqrt{2}}f_{\pi}F^{B\pi}_0(m_\pi^2)
(m_B^2-m_\pi^2)[-c_1 -\xi c_2
+\xi c_3^{cu} +c_4^{cu}\nonumber\\
& +& {3\over 2}(c_7^{cu} +\xi c_8^{cu} - c_9^{cu}
-\xi c_{10}^{cu}) -{1\over 2}(\xi c_9^{cu}
+c_{10}^{cu})\nonumber\\
&+& {2m_\pi^2 \over (m_b-m_d)(2m_d)}
(\xi c_5^{cu} +c_6^{cu} -{1\over 2}(\xi c_7^{cu} + c_8^{cu}))]\;,\nonumber\\
T_d(\pi^+\pi^-) &=& i{G_F\over \sqrt{2}}f_{\pi}F^{B\pi}_0(m_\pi^2)
(m_B^2-m_\pi^2)[\xi c_1 + c_2
+\xi c_3^{cu} +c_4^{cu} + \xi c_9^{cu} + c_{10}^{cu}\nonumber\\
&+& {2m_\pi^2 \over (m_b-m_u)(m_u+m_d)}
(\xi c_5^{cu} +c_6^{cu} +\xi c_7^{cu} + c_8^{cu})]\;,\nonumber\\
T_s(\pi^0 \bar K^0) &=& i{G_F\over \sqrt{2}}\{f_{\pi}F^{BK}_0(m_{\pi}^2)
(m_B^2-m_K^2)
[ c_1 + \xi c_2 -{3\over 2}(c_7^{cu} +\xi c_8^{cu} - c_9^{cu}
-\xi c_{10}^{cu}) ]\nonumber\\
&-&f_K F^{B\pi}_0(m_K^2)(m_B^2-m_\pi^2)[\xi c_3^{cu}+ c_4^{cu} - {1\over 2}
(\xi c_9^{cu}+c_{10}^{cu})\nonumber\\
& +& {2 m_K^2 \over (m_b-m_d)(m_d+m_s)}
(\xi c_5^{cu} +c_6^{cu} -{1\over 2}(\xi c_7^{cu} + c_8^{cu}))]\}\;,
\end{eqnarray}
where $c_i^{cu} = c_i^c-c_i^u$,  and $\xi = 1/N_c$
with $N_c$ being the number of color. The amplitude
$P_{d,s}$ are obtained by setting $c_{1,2} = 0$ and changing
$c_i^{cu}$ to $c_i^{ct} = c^c_i - c^t_i$.
 We have used the following decompositions for the
form factors
\begin{eqnarray}
&<&\pi^+(q)|\bar d \gamma_\mu(1-\gamma_5) u|0> = if_\pi q_\mu\;,
<K^+(q)|\bar d \gamma_\mu(1-\gamma_5) u|0> = if_K q_\mu\;,\nonumber\\
&<&\pi^-(k)|\bar u \gamma_\mu b|\bar B^0(p)> = (k+p)_\mu
F^{B\pi}_1+(m_\pi^2-m_B^2){q_\mu\over q^2}(F^{B\pi}_1(q^2)
-F^{B\pi}_0(q^2))\;,\nonumber\\
&<&K^-(k)|\bar u \gamma_\mu b|\bar B^0(p)> = (k+p)_\mu
F^{BK}_1+(m_\pi^2-m_B^2){q_\mu\over q^2}(F^{BK}_1(q^2)
-F^{BK}_0(q^2))\;.
\end{eqnarray}

It is a well known fact that in order to obtain asymmetry in rates, the decay
amplitudes must contain relative weak and strong rescattering phases.
In our case the weak phases are provided by the phase in
the CKM matrix elements. For the strong rescattering phases, we use the
phases generated at the
quark level with the averaged $q^2 = m_b^2/2$ in eq.(11). The strong
phases generated this way are about $10^o$.
The final results for the asymmetries are given in Fig. 1. In the case of
$\bar B^0\rightarrow \pi^0\pi^0$, we obtain a large asymmetry if $sin\gamma$
is large.
The asymmetry can be as large as 12\%.
The asymmetry for $\bar B^0\rightarrow \pi^+\pi^-$ is smaller by a factor about
3.
In the case for
$\bar B^0\rightarrow \pi^0  K_S$, we also obtain smaller asymmetry.
In a previous paper we have shown\cite{8} that the rate difference
$\Delta(\pi^0\pi^0) = \Gamma(\bar B^0\rightarrow \pi^0\pi^0) -
\Gamma(B^0\rightarrow \pi^0\pi^0)$ is equal to $\Delta(\pi^0 K^0)
= \Gamma( B^0\rightarrow \pi^0 K^0) - \Gamma(\bar B^0\rightarrow \pi^0\bar
K^0)$
in the SU(3) limit. This gives $\Delta(\pi^0\pi^0) = 2\Delta (\pi^0 K_S)$. One
naively expects the asymmetries for both cases to be of the same order.
However,
this is not
the case because the decay amplitude for $\bar B^0\rightarrow \pi^0\pi^0$
is dominated by the tree amplitude which in proportional to $a_2$ and is
therefore
suppressed, thus increasing the asymmetry, while the decay amplitude for
$\bar B^0\rightarrow \pi^0 K_S$
is dominated
by the penguin contributions which is not suppressed, and therefore resulting
in a smaller
asymmetry\cite{dp}. We used two sets of different form factors evaluated in
Ref.\cite{9}
and Ref.\cite{10}.
It is interesting to note that the asymmetries in $\bar B^0 \rightarrow
\pi^0\pi^0$ and $\bar B^0 \rightarrow \pi^+\pi^-$ are
insensitive to the choice of the form factors.

The measurement of asymmetry $Asy$ can also be used in principle to obtain
information
about the weak phase angle $\gamma$ through the use of eq.(8) with
$\delta_w
= \gamma$. There are likely to be errors in $T$ and $P$ evaluation using
factorization approximation, but the ratio $P/T$ is probably more reliable.
The largest source of uncertainty is from the evaluation of $\delta_s$.
$\delta_s$ calculated at quark level is approximately $10^o$ and this
may be  good  to about 30\%. If that is indeed the case, $\gamma$ will be
determined with the same error. Further improvements in the theoretical
treatment of nonleptonic B decays are required for a more defenitive
determination of the weak phase.

We would like to point out that
measurements discussed here will also have great impact on the efforts to
test the SM by measuring the CKM untarity triangle. To measure some of
the phase angles in the unitarity triangle,
it is necessary to measure rate asymmetries in time evolution at asymmetric
colliders
\cite{stone,4,11}.
There are two terms varying with time, one
varys as a cosine function and the other as a sine function. The coefficient
$C_s$
of the sine term contains information about the phase angles in the unitarity
triangle.
If the coefficient $C_c$ (proportional to Asy) of the cosine term
is not much smaller than $C_s$,
like the case for $\bar B^0\rightarrow \pi^0\pi^0$, without
knowing the precise value for $C_c$ the measurement for $C_s$ will be
difficult.
Precise measurements of both coefficients $C_{s,c}$ are required. Although
$C_c$ can also be measured at asymmetric colliders, it is clear that
independent measurements of $C_c$ from a symmetric collider
will provide useful information for determining $C_s$ at an asymmetric
collider.

To conclude, we have shown that contrary to the conventional belief time
integrated asymmetry are measurable in selected final states in the neutral $B$
system at symmetric colliders. These asymmetries are indications of direct CP
violation and would rule out superweak thoeries that have CP violation only in
$B^0-\bar B^0$
mass matrix.
Our factorization approximation calculation
indicates that CP
asymmetry in $\bar B^0\rightarrow \pi^0\pi^0$ can be as large as 12\%. The CP
asymmetries in $\bar B^0\rightarrow \pi^+\pi^-$ and $\bar B^0\rightarrow \pi^0
K_S$ are smaller.

We would like to thank Sheldon Stone for carefully reading the manuscript
and making useful suggestions.

\begin{figure}[htb]
\centerline{ \DESepsf(sym-collider.epsf width 10 cm) }
\smallskip
\caption {The asymmetry as a function of the phase angle $\gamma$.
The horizontal axes are $\gamma$ in degrees. The
solid and dot-dashed lines in Fig. 1a are
for asymmetries in $\bar B^0\rightarrow \pi^0\pi^0$ and
$\bar B^0\rightarrow \pi^+\pi^-$, respectively. The solid and dashed
lines in Fig.1b are for asymmetry in $\bar B^0\rightarrow
\pi^0 K_S$ using
the form factors in Ref.[11] and Ref.[12], respectively.}
\label{gamma}
\end{figure}


\begin{references}
\bibitem{dp}
In the above we have neglected the effects on the asymmetry from the operator
$O_{11}$ which has been shown to have important effects on some
$B$ decays\cite{ref}.
We have carried out a calculation using the
same approximation discussed in Ref.\cite{ref}, we find that the effect on
the asymmetry for $\bar B^0\rightarrow \pi^0\pi^0$ is negligiblely small and
the effect on $\bar B^0\rightarrow \pi^0\bar K^0$ is less than 25\%.
\end{references}
\end{document}